\documentclass[preprintnumbers]{revtex4}
\usepackage{natbib}
\usepackage{amsmath}
\usepackage{amstext}
\usepackage{amssymb}
\usepackage{latexsym}
\usepackage{theorem}
\usepackage{placeins}
\usepackage{moreverb}
\usepackage[amssymb]{SIunits}
\usepackage{xspace}
\usepackage{doublespace}
\usepackage{epsfig}
\usepackage{psfig}
\usepackage{wrapfig}
\usepackage{hyperref}
\usepackage{epstopdf}

\newcommand{\ie}{i.e.{}\xspace}			
\newcommand{\viz}{viz{}\xspace}			
\newcommand{\etc}{etc.\xspace}			
\newcommand{\cf}{cf.{}\xspace}			
\newcommand{\ibid}{\emph{ibid.}}		
\newcommand{\perse}{\emph{per se}\xspace}
\newcommand{\apriori}{\emph{a priori}\xspace}

\newcommand{\Exp}[1]{\langle #1 \rangle}

\newcommand{\proof}{\noindent{\bf Proof.~}}

\newcommand{\discussion}{\noindent{\bf Discussion.~}}

\newcommand{\Qed}{$\mathbf{\Box}$}

\newtheorem{lemma}{Lemma}
\newtheorem{theorem}{Theorem}

\newtheorem{corlemma}{Corollary}[lemma]

\newcommand{\Label}[1]{\ \\ \textbf{#1}}
\newcommand{\LI}{\begin{itemize}}
\newcommand{\IE}{\end{itemize}}
\newcommand{\LN}{\begin{enumerate}}
\newcommand{\NE}{\end{enumerate}}
\newcommand{\LD}{\begin{description}}
\newcommand{\DE}{\end{description}}
\newcommand{\LL}{\begin{list}}
\newcommand{\LE}{\end{list}}

\begin{document}
\title{
	$h$ is classical
	}
\author{V. Guruprasad}
\begin{abstract}

Regardless of number,
standing wave modes are by definition
	noninteracting,
and therefore cannot thermalize
	by themselves.
Doppler shifts due to thermal motions of cavity walls
	provide necessary mixing,
but also preserve
	the amplitudes and phases.
The $\lambda/2$ intervals of the modes thus preserved
	must have equal energy expectations,
		say $\langle \epsilon \rangle$,
		in the resulting equilibrium.
By definition again,
they can be exchanged between modes
	only in whole numbers
and hence	
	only between harmonics.
Each family of harmonic modes is thus self-contained
	and is disjoint from other families
		in such exchanges,
and further,
	can have no more than one mode excited
		at any instant.
The second property identifies
	harmonic families of standing wave modes
		as the harmonic oscillators
	of Planck's theory,
since a family can only bear energy equal to
	exactly one of $\epsilon$, $2\epsilon$, $3\epsilon$, \etc
These two properties further imply that 
	the energy expectation gets averaged only over
		an individual family,
as the equilibrium energy
	$\langle E(\nu) \rangle$
	steadily available at a given mode
		would have contributions
	from its entire harmonic family.
Planck's equations reemerge,
and radiation quantization arises as
	a classical rule
	$
	\langle E(\nu) \rangle
	=
		\langle \epsilon \rangle
		\nu
	$,
	as a mode must contain a whole number of
		exchangeable $\lambda/2$ intervals,
but it only concerns
	equilibrium states.
The result makes Planck constant $h$
	an analogue of Boltzmann's constant $k_B$ 
		for the frequency domain,
and points to
	postulate-free explanations of all aspects of
		quantum and kinetic theories.

\end{abstract}
\maketitle
\newcommand\ArticleType{paper\xspace}
\newcommand\Section[2]{\section{#2}\label{s:#1}}
\newcommand\Endnote[2]{\endnote{#2}#1\xspace}
\Section{intro}{Introduction}

Presented below is a first-ever analysis of
	the equilibrium of standing wave modes themselves
subject to
	irreversible interactions with the environment
		via the thermal motions of confining walls.
The wall interactions are provided by
	the classical Doppler shifts
applicable to all waves and all wavelengths,
	regardless of
		the molecular, atomic or subatomic properties
			of wall matter,
and overcome Boltzmann's argument of 1897
	that Maxwell's equations are symmetric and 
		do not provide irreversibility
	\cite{PhysWebPlanck2000},
obviating Planck's hypothesis of ``elementary disorder'' [\ibid]
	for the determination of
		the equilibrium spectral distribution.
Two basic aspects of
	this dynamic equilibrium
	are established,
as follows.
\LI
\item 
	The half-wavelength ($\lambda/2$) intervals
		of the standing wave modes
	would be
		statistically preserved in number at all times,
	thus providing
		the exact analogues to ``atoms'' in the kinetic theory
			sought by Planck.
	More particularly,
	the $\lambda/2$ intervals would, in effect, be exchanged
		between standing wave modes
	only in whole numbers.

	Doppler shifts preserve
		both the amplitude and the phase profiles of
			reflected wave trains exactly,
	hence the $\lambda/2$ intervals would be atomic
		under this mechanism,
	but the preceding arguments are
		statistical and independent of the precise mechanisms
	responsible for
		the mixing of the mode energies.
	Absorption and emission by matter
		are still necessary for complete relaxation,
	which should include
		splitting and recombination of wave trains
	but irreversibility \perse is adequately provided by
		the wall shifts and
			does not depend on relaxation.
	This seems to resolve another long pending issue
		in the kinetic theory,
	known as the Fermi-Pasta-Ulam (FPU) problem,
	that
		large number simulation models continue to display
			periodicities uncharacteristic of heat
		\cite{FPU1955}
		\cite[\S5.5.1]{TodaKuboSaito1992}
	--
	for real systems,
	the notion of equilibrium invariably includes
		equilibrium with the environment.



\item 
	The combination of relaxation and irreversible mixing
		should drive the confined radiation to a state
	in which
		all of the entities preserved by the mixing,
			\viz the $\lambda/2$ intervals,
		would have equal expectation energies.
	This incidentally means that
		the equalized energy integrals are
			relative to phase
		rather than physical space.
	Further,
	these expectation energies must match
		those of similar atomic entities
			outside the cavity walls,
		and must thus be universal.
	This, then, is the true equivalent of
		the equipartition in kinetic theory
	for radiation.

	Since the number of $\lambda/2$ intervals
		in a standing wave mode
	is proportional to its frequency, 
	Planck's quantization rule
		$E(\nu) = h \nu$
	is directly reproduced classically
		in this state of dynamic equilibrium.
	Two further observations also follow
		from the definition of standing wave modes,
	\viz
	that the $\lambda/2$ intervals can be exchanged
		only between harmonically related modes,
	and that every such exchange would destroy
		the excitation in the originating mode
		while populating
			the destination mode.
	These further observations suffice,
		as will be shown,
	to identify Planck's oscillators
	with real families of harmonically related modes,
		and to reproduce Planck's law \emph{classically}.
\IE

The present result demolishes
	a century old conclusion,
based on
	an analysis by Rayleigh and Jeans
and informally reinforced by
	Einstein's photoelectricity theory,
that quantization is \emph{absolutely underivable}
		from classical physics
	(\cf \cite[I-41-2]{Feynman}).
Rayleigh and Jeans attributed
	the equipartition energy of $k_B T/2$,
		established for particulate entites
			in the kinetic theory,
	to individual standing wave modes,
and this famously led to
	the problem of ``ultraviolet divergence''.
However,
their more basic error of attributing
	the \emph{entire} time-domain equipartition energy
to \emph{each} Fourier component in the same equilibrium,
	has gone unnoticed
--
this subsumes
	the Rayleigh-Jeans divergence
and questions
	the precise use of Fourier decomposition 
		in the subsequent quantum theory. 
Persisting speculations like
	``Hamilton had no basis to assume that
		$h$ was anything but zero''
	\cite[\S10-8]{Goldstein},
are a case in point,
since Bohr's Correspondence Principle theory
	[\ibid]
	makes $h$ the scale factor
		for the frequency domain
--
and $h = 0$, 
	the only logically inadmissible value!

More generally,
individual Fourier components are notionally identified with
	particles in current physics,
explaining, in hindsight,
	why the Boltzmann statistics,
		which leads to
		the time domain equipartition of $k_B T/2$,
has to be replaced in quantum theory
	by the boson or fermion statistics
applicable specifically, and only, to
	the Fourier components.
This association of Fourier components
	with particulate behaviour
is only partly explained by
	Hamilton-Jacobi theory and the Correspondence Principle,
which leave
	the very nature of quantum wavefunctions,
	the mechanism of ``wavefunction collapse'',
and so on,
	to as yet speculative interpretations.
A complete theory must include
	a very general treatment of
		the physical processes of observation
to actually explain
	the origin of quantization,
\ie mundanely,
	without indirectly reintroducing equivalent postulates.
Such a theory is mandated and indeed made possible
	by the present result
as the separation of
	irreversibility from relaxation
admits irreversible interactions with
	\emph{the Hamilton-Jacobi standing wave components of
		the inequilibrium states of
			the observers and their instruments}
	that represent information
	(\cf \cite{Landauer1961a,Landauer1961b}).
This would suffice to fundamentally ensure that
	the standing wave mode $\lambda/2$ intervals
also necessarily characterize
	\emph{all} observations by \emph{all physical} observers,
thus finally proving
	the universality of quantization
		on formal classical grounds,
limiting
	its applicability to states of thermalization
and reducing
	the probabilistic nature of wavefunctions to heat.

This broader picture would be incomplete without also noting
	why a classical foundation is necessary, and sufficient,
		for a fundamental understanding of nature,
apparently contradicting all of the 20th century wisdom 
	viewing subatomic elementary particles
		as the ultimate constituents of matter.
The necessity arises from the fact that
	real observers and processes of observation
		must be macroscopic by definition,
	and therefore classical.
More particularly,
the observer's physical states
	representative of observed data
must be also macroscopic
	for the same reason,
as well as stationary
	to be representative of knowledge.
The sufficiency follows from the fact that
	the macroscopic observer interface can have 
		at most a finite precision,
hence any finite combination of observer states
	necessarily represents at best
		a continuum of possible configurations of nature
	that cannot be distinguished
		by those states.
In other words,
\emph{the finiteness of real observers
	prevents them from precisely observing and representing
		anything but continuous fields},
equivalent to
	a finite combination of Fourier terms.
Moreover,
the inherently macroscopic nature of observers
	makes it moot
whether the elementary particles
	are as fundamental as currently thought,
or merely reflections of
	the energy being thrown into the observational processes
		in high energy research.
The Hamilton-Jacobi standing wave modes of the observer
	suffice as a basis for representing observable patterns
		at the observer interface,
and would be fundamental in terms of
	reducibility of the laws of physics to
		invariants of these patterns,
providing
	a formal foundation for reasoning in physics
		analogous to the Turing machine
	in computational science.
Classical physics happens to be consistent with this foundation,
to the extent that it relies on definitions
	rather than postulates
	\Endnote{.}{
	Reducibility of
		the constancy of the speed of light postulates
		of special relativity
	is proved in the appendix of
		\cite{Prasad2000c}.
	Current ideas
		that quantum mechanics is irreducible
		and \emph{more} fundamental
	unfortunately place
		modern physics on a speculative,
			almost theological foundation.
	The power of physics lies not in
		exotic leaps of faith,
	but in mundane reasoning
		to understand, describe and
		otherwise deal with the same empirical data.
	The present result should thus be seen as
	advocating and enabling
		a rigorous understanding of quantum physics,
	rather than
		merely questioning its foundation.
	}

Correspondingly,
the particulate approach has led to at least
	two fundamental limitations in modern physics.
First,
it has obfuscated
	the very consideration of successive diffractions,
\ie
	continued bending of already diffracted
		portions of wavefronts
	due to further diffraction by
		other obstructions encountered by these portions.
Successive diffractions are well understood
	in the form of surface effects,
for example,
	in the known capture of passing wavefront energy
		by a sphere virtually forever,
but past considerations of diffraction in
	astronomy and quantum field theories
are limited to Fresnel or Fraunhofer approximations,
	for dust attenuation
		\cite[pp149-153]{Spitzer1978}
	and scattering
		\cite[pp149,158]{Weinberg1995},
		respectively.
These approximations only describe
	total deflections of less than $\pi/2$,
whereas the spherical capture alone suffices
	to change the fundamental wave propagation law
from
	the currently used $r^{-2}$ law for free space
to
	a weakly attentuated form $e^{-\sigma r}r^{-2}$
for both
	the real, matter-filled universe
and
	inside dense matter like the sun's interior.
Its contributions to
	Olbers' paradox,
	galactic dark matter
and
	neutrino oscillations
		remain to be verified.
A second, more basic wave effect,
which was also previously unrecognized
	but will be hopefully soon verified on laboratory scale, 
is a Doppler-like scaling of observed frequencies,
but in proportion to source distances
	like Hubble's law
	\cite{Hubble1929,Prasad2005a,Prasad2005b},
and driven solely by temporal variation
	of the observing instrument and thus its states
	\Endnote{.}{
	This second effect had been suspected
		as far back as 1995,
	leading to informal predictions of
		the cosmological acceleration.
	The pattern of variations in the Pioneer 10 anomaly
		\cite{Turyshev1999,Anderson2002}
	was found to be consistent with
		mundane mechanical creep
	under solar tidal forces and
		the centrifugal force of the spacecraft spin
		\cite{Prasad2006USPpubA}.
	Empirical data ranging from
		Type IA supernovae (the cosmological acceleration)
	to the ``expanding earth'' patterns,
		including GPS station coordinates data,
	seem similarly consistent with
		creep in our instruments
			under earth's gravitational and tidal stresses,
	in combination with
		an unanticipated error systematically introduced
			by our calibration procedures
		[\ibid].
	The diffractive effect can also account for
		the radiation background,
	and together,
		they question relativistic cosmology theories
	on grounds unprecedented in
		generality and rigour.
	} 

The key contributions are thus 
a more precise thermodynamic treatment of
	confined radiation;
the correct principle of equipartition
	for the frequency domain;
separation of irreversibility from relaxation solving
	the FPU problem and improving over Boltzmann's ideas;
and
a principle of reality governing
	every observer of relevance to physics,
	making it necessary to consider
		its representative states
	and leading to
		universal quantization as explained.
The rest of this \ArticleType is concerned with 
	the radiation equilibrium,
and with proving the following theorem
	representing its core result.

\begin{theorem}[Classical equivalence of quantum mechanics]\label{t:equiv} 
	Planck's law implies and is implied by
		a classical equilibrium of
	standing wave modes of radiation
		with their environment.
\end{theorem} 


The forward implication of a classical equilibrium
	is proved first
		in Section \ref{s:implied},
both to establish 
	the consistency of the classical derivation
		to be given in Section \ref{s:proof},
and to provide needed insight into
	the nature of travelling waves.
The main result is of the course
	the derivation itself,
	in Section \ref{s:proof},
proving the reverse implication that
	the classical equilibrium of standing wave modes 
		yields Planck's law,
with no postulate
	that might imply inherent quantization.
Appendix \ref{a:hj} explains
	the Correspondence Principle issue in detail.

\Section{implied}{Quantization implies a classical equipartition}


The forward implication of Theorem~\ref{t:equiv},
that Planck's law in fact implies
	a classical equipartition over
		$\lambda/2$ intervals of standing wave modes,
is proved below
	as a slightly more general lemma.


\begin{lemma}[Spectral equipartition]\label{l:eqphase} 
	For confined radiation in thermal equilibrium,
		Planck's quantization rule,
	\begin{equation} \label{e:ehf}
		E(\nu) = h \nu ,
	\end{equation}
	implies uniform distribution of energy between
		full wavelength intervals
			of arbitrary periodic waveforms,
	and between
		half-wavelength intervals of sinusoidal waves.
\end{lemma}

	\begin{wrapfigure}[12]{r}{2in}
		\centering
		\psfig{file=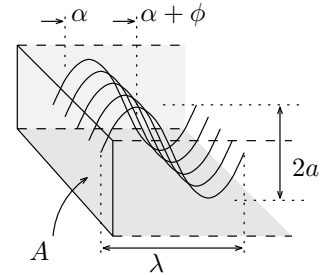}
		\caption{The integral $U(\phi, \alpha, \sin)$}
		\label{f:Sinewaves}
	\end{wrapfigure}
\proof
	As the wave travels through
		any cross-sectional plane in space,
	its phase at the plane
		will vary over a full cycle.
	Hence,
	if we integrate over
		a very small phase interval $\phi \lambda/2\pi$,
			where $\phi \ll 2 \pi$,
	the instantaneous total energy
		in the spatial interval starting at the plane
	will vary between
		$0$ and $a^2 A \, \mathcal{F}(f, \phi)$,
	where
		$A$ is the cross-sectional area chosen
	and
		$\mathcal{F}(f, \phi)$ is a shape factor
			that depends on the waveform $f$,
	with $\mathcal{F}(\sin, \pi) = 1/2$.

	For the identified interval lengths,
	\viz $\lambda$ intervals ($\phi = 2\pi$) in general
		and $\lambda/2$ intervals ($\phi = \pi$)
			for sinusoids,
	the total wave energy in such an interval
		cannot change as the wave passes.
	This is proved as a lemma in
		Appendix \ref{a:lobes}.
	For the sinusoidal intervals,
	the energy integral
		\emph{with respect to phase}
	is
		(see Fig.~\ref{f:Sinewaves})
	\begin{equation} \label{e:phlobe}
		U(\pi, \alpha, \sin)
		\equiv
			A
			\int_{\theta = \alpha}^{\alpha + \pi}
				a^2 \sin^2 (\theta)
				\; d\theta
		\equiv
			A
			\int_{0}^{\pi}
				a^2 \sin^2 (\theta + \alpha)
				\; d\theta
		=
		A \cdot \pi a^2 / 2
	\end{equation}
	and occupies
		a length of $\lambda / 2$.
	It represents an energy density
		$
		u \equiv
			U(\pi, \alpha, \sin) / (A \, \lambda / 2)
		=
			\pi a^2 / \lambda
		\equiv
			\pi a^2 \nu / c
		.
		$

\begin{wrapfigure}[11]{l}{2in}
		\centering
		\psfig{file=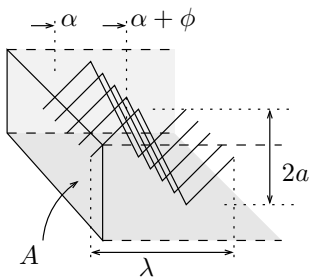}
		\caption{The integral $U(\phi, \alpha, f)$}
		\label{f:Otherwaves}
	\end{wrapfigure}

	The length of
		an interval spanning a phase difference of $\phi$ is
		$
		\lambda \cdot \phi / 2 \pi \equiv c \phi / 2 \pi \nu
		$,
	where
		$\lambda$ is the wavelength,
		$\nu$, its frequency
	and
		$c$ denotes the speed.
	If the internal dimension of the cavity
		along the direction of the wave
	is $L$,
	there would be
		$L \cdot 2 \pi \nu / c \phi$ such intervals,
	and more particularly
		$2 \nu L / c$ of $\lambda/2$ intervals.
	The total wave energy is then
	\begin{equation} \label{e:wave}
		E(\nu)
		=
		(A \lambda / 2) \cdot u \cdot 2 \nu L / c
		=
		a^2 \pi \nu V / c
		,
	\end{equation}
	where $V \equiv A L$ is the volume,
		for a rectangular cavity.
	When the cross-section is uneven,
		$A$ would need to be integrated along $L$,
	but the overall result
		would be clearly preserved.
	Dividing equation (\ref{e:ehf}) by
		this total wave energy yields
	\begin{equation} \label {e:constwave}
		a^2
		= h c / \pi V
		,
	\end{equation}
		\emph{which is independent of the frequency}.
	Thus, the quantization rule,
		equation (\ref{e:ehf}),
	means that
		sinusoidal waves will all have
			the same mean amplitude $|a|$,
	\ie a uniform distribution of energy
		among $\lambda/2$ intervals
			regardless of $\lambda$.
	The generalization to
		$\lambda$ intervals of
			an arbitrary periodic waveform
		(Fig.~\ref{f:Otherwaves})
	follows from
	the Fourier series decomposition of such a waveform
		into a whole number of wavelength intervals of
			its sinusoidal components.
	By a lemma proved in Appendix \ref{a:lobes},
		which is applicable to these components,
	the phase offset of
		the full wavelength interval of the parent waveform
	cannot matter
		with respect to its energy.

	The Fourier decomposition is equal to the sum of
		an infinite series of sinusoidal waves
	with nodes coinciding with
		the end points of its period,
	as
	\begin{equation} \label{e:fseries}
		f(x) =
		\sum_{n = 0}^{\infty}
			F_n
			\,
			e^{i k_n x}
	\quad
	\text{where~}
		k_n = 2 \pi / \lambda_n
	\text{~and~}
		F_n =
		\int_{x = 0}^{\lambda}
			f(x)
			\,
			e^{i k_n x}
		dx
		.
	\end{equation}
	Equation (\ref{e:phlobe}) must hold for
		each of these sinusoidal components,
	provided that we refer
		the corresponding phase variables
			$\phi_i$ and $\alpha_i$
		for each component
			to the parent waveform.
	The energy of a full wavelength interval of
		the parent is then
	\begin{equation} \label{e:sumseries}
		U^*(\phi, \alpha, f) =
		\sum_{n = 0}^{\infty}
			F_n
			\,
			U (n \phi , n \alpha , \sin )
			.
	\end{equation}
	Equality of the energy sums $U^*(2\pi, \alpha, f)$
	is then assured regardless of
		$\alpha \in [0, 2\pi)$,
	as the condition individually holds
		for each of the sinusoidal components
	independently of their starting phases,
		as remarked.
\Qed

\vspace{5pt} \discussion 
This is a more general result than needed
	for Theorem \ref{t:equiv},
since it holds for
	$\lambda$-intervals of all waveforms,
instead of merely
	the sinusoidal components constituting
		the standing wave modes.
It suggests
	a deeper significance of Planck's discovery
which has hitherto escaped attention,
\viz
that (half-)wavelength intervals of unit cross-section behave
	like particulate bearers of energy,
similar to
	the molecules of a gas in the kinetic theory.
The equipartition is different
	from that of kinetic theory,
however,
since we have
	\emph{equal amplitude expectations},
instead of
	the \emph{equal energy expectations}.
A na\"ive application of
	the classical equipartition
would be based on
	the energy integral \emph{over space},
\begin{equation}\label{e:classicalenergy}
	V(x_1, x_0, \sin) \equiv
		\int_{x_0}^{x}
			a^2 \sin^2(kx')
			\; dx'
	\equiv
		\frac{a^2}{k}
		\int_{\theta = kx_0}^{kx}
			\sin^2(\theta)
			\; d\theta
	\equiv
		\frac{a^2}{k}
		\int_{kx' = kx_0}^{kx}
			\sin^2(\theta + kx_0)
			\; d\theta
	\equiv
		\frac{a^2}{k}
		\; U(kx, kx_0, \sin)
	.
\end{equation}
However,
equation (\ref{e:classicalenergy}) is merely
	the energy density of a wave by itself.
It has no bearing on
	the thermal equilibrium
as the classical principle of superposition of
	electromagnetic fields
implies
	a complete absence of dynamical interaction
		between their waves.
This is in effect why Planck was forced to stipulate
	his principle of ``elementary disorder'',
which was interpreted in the Rayleigh-Jeans analysis
	in terms of absorption and emission by atomic matter
	\cite[I-41-2]{Feynman}.
The equal energy approach is inherently flawed,
however,
	when applied to standing wave modes
since the latter constitute
	individual Fourier components of radiation,
rather than
	independent, full-fledged dynamical entities,
		as remarked.
In particular,
	as the frequency domain is infinite,
the idea inherently implied
	infinite energy.

Lemma \ref{l:eqphase} favours Doppler shifts
	as the driving mechanism for thermalization,
since they are the only mechanism for
	energy transfer between standing wave modes
guaranteed to preserve
	amplitudes and phase intervals.


\Section{proof}{Classical equilibrium yields quantization} 


The reverse implication of
	Planck's law and the radiation quantization
from
	the classical equilibrium of the standing wave modes
requires showing first that 
	the standing wave modes are as such
		a complete set of interacting entities
for describing 
	the radiation equilibrium,
and then showing that
	Planck's law results from
		their classical equilibrium.

The equilibrium is considered to be
	of the modes themselves,
which closely relates to
	the Bose-Einstein derivation
		and second quantization picture,
instead of
	a mere balance of power flows
		in interactions with matter
	as considered in
		the Rayleigh-Jeans analysis.
The key difference from both
	Planck's and the Bose-Einstein derivations
is, of course, that
	these interacting entities are real and mundane,
		in existence as well as properties,
rather than
	postulated entities with hypothesized behaviour
that happened to yield
	the same spectral distribution.

This more substantive role of standing wave modes,
of actually characterizing
	the equilibrium distribution of radiant energy
		by themselves
with no more than
	perfectly classical interactions,
had thus remained unobvious,
likely thanks to distraction 
	by the notion of quantization.
It is accordingly established first
	as a lemma below.
The rest of the proof of Theorem \ref{t:equiv}
follows from Lemma \ref{l:boltzmann},
	deriving Planck's law from
		this classical equilibrium.
The universal nature of $h$ follows from
	the form of Planck's law,
		just as in traditional theory,
and is included as a corollary
	for reference.


\begin{lemma}[Sufficiency of standing wave modes]\label{l:set} 
	Standing wave modes constitute a complete set for representing
		the equilibrium energy distribution of confined radiation.
\end{lemma}

	\begin{wrapfigure}[5]{r}{2in}
		\centering
		\psfig{file=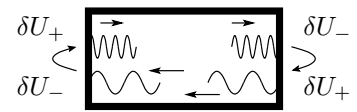}
		\caption{Mixing at the walls}
		\label{f:Mixing}
	\end{wrapfigure}
\proof
	The total energy $U$ in the cavity
		is instantaneously comprises
	three parts:
	\LI
	\item
		a portion $U_0$ contained in the standing wave modes,
	\item
		a portion $U_-$ contained in travelling waves
		that happen to be losing energy to other frequencies
			due to the wall Doppler shifts
			(Fig.~\ref{f:Mixing}),
		and
	\item
		a similar travelling wave portion $U_+$ gaining energy
			from other frequencies.
	\IE
	We have $dU_0/dt = 0$ by definition, hence
	statistical constancy of the total energy in equilibrium
		means that $dU_-/dt = - dU_+/dt$.
	Any occurrence of 
		$U_-$ and $U_+$ components signifies
			mixing of wave energies across frequencies.
	The mixing must occur due to
		absorption and emission by
	matter
		comprising the confining walls
	or
		contained within the cavity,
	as well as by Doppler shifts
		from the thermal motions of the walls.
	The wall motions are implied by
		the very premise of equilibrium
			even in Planck's theory,
	because for equilibrium at a temperature $T$,
		the cavity must be kept in thermal contact with
			a reservoir at $T$.

	As remarked in the Introduction,
	the wall Doppler shifts cannot account for
		the relaxation of
			any \apriori excited standing wave modes
	as Doppler shifts preserve
		both amplitudes and phase extents of
			reflected wave trains.
	They do suffice to introduce irreversibility,
	however, and to effect mixing of mode energies
		at a potentially greater rate
		limited only by
			the wave speed within the cavity,
	separately from the relaxation due to
		absorption and emission by matter.
	The precise rates are not important
		for the equilibrium distribution,
	so long as
		they are nonzero.

	Consider first the hypothetical ideal conditions of
		perfectly rigid, non-interacting walls
	that neither absorb nor emit energy,
		and must therefore perfectly reflect
			all incident radiant energy.
	This was implicitly assumed in
		Planck's theory and also in
			the Rayleigh-Jeans analysis,
	as both ignored
		the possible thermalizing effects of wall motions.
	Incidentally,
	this has also been a subtle limitation 
		in the kinetic theory of gases,
	where the notion of entropy
		given by Boltzmann's $H$-theorem
	has been historically attributed to
		statistical probabilities of large numbers alone,
	but this has been shown by simulations to be inadequate,
		as mentioned.

	In the present context of confined radiation,
	for every incident travelling wave $w_{(-)}$,
	there must then be
		an outgoing reflected travelling wave $w_{(+)}$
		of exactly the same frequency and amplitude,
	since there would be no shift of frequency
		and the instantaneous power $dU(w_-)/dt$
			of the incident wave
		must be exactly balanced by
			the power $dU(w_+)/dt$ of the reflected wave.
	The incident and reflected waves
		therefore form standing waves,
	since the condition amounts to 
		$dU(w_-)/dt + dU(w_+)/dt = 0$
			for each $w_{(-)}$.
	These idealized conditions thus permit
		only standing waves,
	hence the standing wave modes necessarily comprise 
		the complete energy spectrum of the radiation.

	In the real world scenario
		where the walls are not perfectly reflecting,
	that is,
		where atomic interactions and thermal Doppler shifts
			are involved,
	there would be
		a mean outflow $dU_-/dt$ of energy 
			going into the walls and atoms therein,
		and a statistically equal inflow $dU_+/dt$
			entering the cavity,
	for the statistical constancy of
		the total energy $U$.

	We cannot assume \apriori that
		the instantaneous values of $dU_-/dt$ and $dU_+/dt$
			will also match at every frequency,
	as imperfect reflectivity means that
		some frequencies will likely be gaining energy,
	but the gain must eventually occur at the expense of others
		for the statistical constancy of $U$
	over
		an extended period of observation.
	Instantaneously,
	there will be travelling waves going into the walls
		and losing energy at some frequencies
	as the outbound waves happen to be absent or weaker
		at these frequencies,
	and likewise,
	there will be waves emanating from
		the walls and the atoms therein
			at other frequencies with
		weaker or no corresponding inbound waves.
	Over a sufficiently long interval of time,
	there should be no net gain or loss
		at any individual frequency $\nu$.

	Since the expectation value $E(\nu) = \Exp{U(\nu)}$
	is defined by
		averaging over all such waves over
			a long enough period of observation
		(ideally infinite),
	it would be quantitatively equal to the average computed
		when no gain or loss occurred at all
			at any frequency $\nu$
		over that same interval.
	This is the same as assuming
		that all of the energy at frequency $\nu$
			had been contained solely in standing wave modes.
\Qed 

\Label{Classical equipartition of radiation.}
According to classical statistical mechanics,
in the state of equilibrium of
	any system of interacting entities,
		such as gas molecules in the kinetic theory,
there is
	equal probability of finding
		any individual entity at energy $u'$,
given by 
	the probability density function
\begin{equation} \label{e:UBoltz}
	p
	\equiv
		p ( u' )
	\propto
		e^{-u' / k_B T}
	.
\end{equation}
We would expect equation (\ref{e:UBoltz}) to be
	identically applicable for the radiation equilibrium
in terms of
	any set of energy-bearing components of the radiation
		in the cavity
that, like the molecules of a gas,
	will ordinarily be neither created nor destroyed
but can possess
	different energies.

The premise of equilibrium at a temperature $T > 0$
as such implies
	equilibrium with the immediate environment
		also at $T$,
and this implies
	involvement of wall Doppler shifts
as a key, classical mechanism of irreversibility,
	as mentioned in the Introduction.
It has the remarkable property of also conserving
	the number of $\lambda/2$ intervals
		of both travelling and standing waves,
but according to the preceding lemma,
	the latter contain all of the radiant energy
		in equilibrium.
To be precise,
the Doppler shifts only suffice to translate
	any standing wave excitation
		to various other frequencies
	in a random sequence.
Absorption and emission by matter
	is necessary for dispersing
		individual excitations to
	multiple, concurrent wave trains,
and would break
	the conservation of $\lambda/2$ intervals
		during relaxation.
In equilibrium, however,
these relaxation transformations
	would have already arrived at a steady state,
so that the conservation of $\lambda/2$ intervals
	would be still assured in a statistical sense.
Equation (\ref{e:UBoltz})
should be therefore applicable to
	the sinusoidal $\lambda/2$ intervals of
		unit cross-sectional area
for which
\begin{equation} \label{e:U}
	u' \sim
	u_\pi \equiv U(\pi, \alpha, \sin) / A
	=
		\pi a^2 / 2
	.
\end{equation}
It follows that in the classical equipartition,
all $\lambda/2$ intervals
	would exhibit the same expectation energy $u'$
		with the probability $p(u')$ 
	given by equation (\ref{e:UBoltz}).
This suffices to yield Planck's law,
	as asserted by the following lemma.


\begin{lemma}[Classical distribution] \label{l:boltzmann} 
	A Boltzmann distribution of the half-wavelength intervals of
		sinusoidal standing wave modes
	yields Planck's law.
\end{lemma}
\proof
	By the preceding discussion,
	this result represents
		radiation equilibrium in general,
	inclusive of all contributing mechanisms,
	even relaxation processes
		that do not preserve the number of
			the $\lambda/2$ intervals.
	Only rectangular cavities need to be considered,
	as the expectation energy distribution of
		an irregular cavity in equilibrium
			with a rectangular cavity
	must match
		the latter's distribution at each frequency,
	by the principle of detailed balance.

	The two component aspects of most interest would be
		the reproduction of Planck's quantization rule,
			equation (\ref{e:ehf}),
	and
		the identification of classical radiative entities
	exhibiting the very properties
		required by Planck's hypothesis of
			quantized harmonic oscillators.
	The first is readily obtained by applying
		Section \ref{s:implied} in reverse
	to the equipartition just described,
	since
		a standing wave mode of frequency $\nu$
		must geometrically comprise
			$2 \nu L / c$ of $\lambda/2$ intervals
		(equation \ref{e:wave}),
	so that the instantaneous mode energy,
		for unit cross-sectional area,
	is given by
	\begin{equation} \label{e:emode}
		E(\nu)
		=
			\frac{2 \nu L u' }{ c }
		\propto
			\nu u'
		,
	\end{equation}
	though with the Boltzmann probability
	\begin{equation} \label{e:pmode}
		p(\nu)
		=
			p^{2 \nu L / c}
		,
	\end{equation}
	given by equation (\ref{e:UBoltz}) applicable to
		each $\lambda/2$ interval within the mode.

	Planck's harmonic oscillators are given,
		as stated in the Introduction,
	by \emph{harmonic families},
	\ie
	families (or sets) of harmonically related standing wave modes,
		$H\{\nu_0\} \sim \{\nu_0,~2\nu_0,~3\nu_0,~...\}$,
	for
		any chosen mode of frequency $\nu_0$.
	We need only
		the following two properties
	for this identification to hold:
	\LI
	\item
		a harmonic oscillator may oscillate at any of
			a discrete set of harmonically related frequencies
		and no other; and

	\item
		a harmonic oscillator can be at only one frequency $\nu$,
		or equivalently,
			at one energy level $E = h \nu$,
		at any instant.
	\IE
	The first property already follows from
		the definition of the harmonic families.
	The second property is not so trivial,
	as multiple members of a harmonic family
		ordinarily can be in excited states
			with various energy levels
	simultaneously.
	However,
	the equilibrium state 
		statistically preserves $\lambda/2$ intervals,
	hence the equilibrium interactions are equivalent to
		exchanges only of whole numbers of
			$\lambda/2$ intervals
		between standing wave modes.
	Further,
	the exchanges happen to be impossible,
		by the geometrical nature of standing wave modes, 
	unless they are strictly between
		members of the same harmonic family.
	A single interval exchange between two modes,
		possessing $m$ and $n$ $\lambda/2$-intervals,
			respectively,
	means the pair of transitions
	\begin{equation*}
	\left.
		\begin{aligned}{}
			[m] & \rightarrow & [m + 1]
		\\
			[n] & \rightarrow & [n - 1]
		\end{aligned}
	\;
	\right\}
	\quad
	\text{or}
	\quad
	\left\{
	\;
		\begin{aligned}{}
			[m] & \rightarrow & [m - 1]
		\\
			[n] & \rightarrow & [n + 1]
		\end{aligned}
	\right.
	,
	\end{equation*}
	where $[r]$ denotes
		a standing wave mode of $r$ whole $\lambda/2$ intervals,
	but each such interacting pair $[m]$ and $[n]$ represent 
		$m$-th and $n$-th harmonics of a fundamental mode
			that would be denoted $[1]$.
	\emph{The state of equilibrium thus represents
		a disjoint partitioning of both
			standing wave modes and their interactions
				into harmonic families}.
	The second property,
	that each family acts as if it has no more than
		one active member at any time,
	follows upon noting that
		each of these exchange interactions logically destroys
			the original pair of standing waves.
	Equation (\ref{e:pmode}) implies
		$
		p([n]) = p^n
		.$

	Equation (\ref{e:emode}) means simply that
	measurements of
		a standing wave mode of frequency $\nu$
	will yield
		the energy value $E(\nu)$
	only with the probability $p(\nu)$
		given by equation (\ref{e:pmode}).
	The equilibrium energy spectrum requires
		the expectation value $\Exp{E(\nu)}$,
	which has to be computed by
		averaging the energies in all modes,
			weighted by their probabilities,
	that can interact
		with the modes at frequency $\nu$.
	By the preceding arguments of
		the equilibrium state and Doppler mixing,
	this means
		all modes of frequency $\nu$ and their harmonics,
	since any of the harmonic waves can contribute
		one or more of their $\lambda/2$ intervals
			to the measured modes.

	The averaging is only necessary
		over a single harmonic family,
	because other harmonic families,
		belonging to other polarizations,
			directions or propagation paths,
		are effectively noninteracting,
	and will independently
		yield a similar distribution.
	Their contribution can be accounted for 
		by incorporating a density of modes factor,
			which amounts to $8 \pi \nu^2 / c^3$,
	as reproduced in Appendix \ref{a:modes}
		from basic texts for reference.
	Further,
	the averaging needs to be only over
		the harmonic family of
		the immediatedly measured frequency $\nu$,
	and excluding even its subharmonics,
	because while subharmonic standing waves
		can contribute power at the measured frequency
			by supplying $\lambda/2$ intervals,
	the entire measured mode cannot be replenished from
		one of its subharmonics.
	This corresponds to
	the equivalent equilibrium distribution of
		the standing wave modes
	instead of
		simply the $\lambda/2$ intervals,
	and in this picture,
		the full set of
	$2 \nu L / c$ $\lambda/2$-intervals of a mode
		must be exchanged as a whole.
	The sum of energy densities in
		the harmonic family of $\nu$
	is
	\begin{equation} \label{e:Ufamily}
	E(H\{\nu\}) =
		u_\pi \times [ \nu \cdot p^{\nu}
		+ 2 \nu \cdot p^{2 \nu}
		+ 3 \nu \cdot p^{3 \nu}
		+ ... ]
		=
		\frac	{ u_\pi \nu \cdot p^{\nu} }
			{ ( 1 - p^{\nu})^2 }
		,
	\end{equation}
	and this is spread over
		a total of
	\begin{equation} \label{e:nfamily}
	N(H\{\nu\}) =
		1 + p^{\nu} + p^{2 \nu} + p^{3 \nu} + ...
		=
		\frac{1}{ 1 - p^{\nu}}
	\end{equation}
		modes in the harmonic family,
	hence the expectation energy density
		at $\nu$,
	neglecting the density of modes, is
  	\begin{equation} \label{e:Uexpect}
	\Exp{u_\pi ( {\nu} )}
		\equiv
			\frac{E(H\{\nu\})}{N(H\{\nu\})}
  		=
  			\frac{ u_\pi \nu }{ p^{-\nu} - 1 }
  		=
  			\frac{ u_\pi \nu }
			     { e^{u_\pi \nu / k_B T } - 1 }
  		.
  	\end{equation}
	With the density of modes,
  	the spectral energy density becomes
  	\begin{equation} \label{e:Planck}
		\Exp{E(\nu, T) \, d \nu} 
  		=
  			\frac{8 \pi \nu^2}{c^3}
  			\frac{ u_\pi \nu \, d \nu }
  				{ e^{u_\pi \nu / k_B T } - 1 }
  		.
  	\end{equation}
  	This is Planck's law,
  	but with the $\lambda/2$ interval energies
		$u_\pi \sim u(\nu)$ instead of $h$,
	and including
		the same intermediate equations
  		(\ref{e:Ufamily})-(\ref{e:Uexpect})
	as in Planck's derivation
		\cite{Planck1901}.
	The form of the quantum law is thus reproduced,
	which was hitherto thought to be
		inherently quantum mechanical
	and impossible to derive
		without the quantum assumptions of harmonic oscillators
			or boson statistics.
	The constancy of
		the equilibrium value of $u_\pi$
	is established as Corollary \ref{cl:h} below,
		also without invoking
			any nonclassical reasoning,
	so that
		$u_\pi$ may be replaced with $h$
			in equation (\ref{e:Planck})
	without diminishing
		the strictly classical character of the present result.
\Qed

\begin{corlemma}[Constancy of wavelength interval energies]\label{cl:h} 
	The equilibrium $u_\pi$ is a thermodynamic constant.
\end{corlemma}
\proof
	By equation (\ref{e:Planck}),
	the equilibrium $u_\pi$ does not depend on
		the total energy, temperature or the geometry
			of a cavity.
	If the cavity is enlarged,
	energy must be added to occupy
		the additional half-wavelength intervals
	that become available from the increased volume,
		in order to keep its temperature steady.
	To raise the temperature alone,
		without changing its geometry,
	energy must be added once again to occupy
		the higher frequency modes,
	which would have larger numbers of
		standing wave $\lambda/2$ intervals,
	as required by equation (\ref{e:Planck})
		for raising the temperature.

	For the further conclusion that the equilibrium $u_\pi$
		is the same for all cavities in equilibrium,
	we must consider the state of equilibrium
		between pairs of cavities
			thermally coupled to each other,
	via a channel for exchanging travelling waves
		or by contact between their walls,
	permitting energy exchange indirectly
		through the wall Doppler shifts.
	Then,
	if the cavities are measured to be
		initially at the same temperature,
	subsequent thermal interaction between them
		should cause no change in their total energies
			or their spectra,
	as such a change would mean that
		the net entropy of the two cavities
			increased upon contact,
	and hence that
		they could not have been at the same temperature.
	Then at each frequency $\nu$,
		their spectral energy densities
		$
		\Exp{E(\nu, T)}
		$
	should match as well,
		as a requirement for detailed balance.
	This would be impossible unless
		$u_\pi$ is the same in both,
	given that $T$ also is the same
		and all of the remaining factors in
			equation (\ref{e:Planck})
		are constants.
	The equilibrium value of the $\lambda/2$ interval energy,
		$u_\pi$,
	is therefore a universal constant,
	signifying
		a thermodynamic scale factor of energy
			for the frequency domain.
\Qed

\Label{Proof of Theorem \ref{t:equiv}.}
	The present analysis has established both that
		Planck's law implies a classical equipartition
		(Lemma \ref{l:eqphase}),
	and that
		the classical equipartition of confined radiation
			given by the irreversibility of
				wall Doppler shifts
		yields Planck's law
		(Lemma \ref{l:boltzmann} and Corollary \ref{cl:h}),
	as required by
		the statement of Theorem \ref{t:equiv}.
\Qed



\Section{concl}{Conclusion} 

A fundamental if hitherto unanticipated
	unification of classical and quantum theories
		has been established above,
by rigorous, detailed analysis,
	evidently for the first time,
of
	the (classical) energies of standing wave modes
		of radiation in a cavity
and of
	their dynamical (classical) equilibrium
under 
	mixing by Doppler shifts due to
		the thermal motions of the cavity walls.
This at first may not seem too surprising
	to some readers
as the radiation equilibrium concerns standing wave modes,
	which are inherently discrete
--
the discreteness of standing wave modes
	is however spatial,
and the difficulty all along has been that
	quantization concerns discretization of energy,
which is at best temporal and is not obvious
	from the spatial discreteness of modes.
Many readers will presumably also contend
	that the present result,
		being specific to the radiation spectrum,
cannot possibly imply
	that classical physics subsumes all of quantum theory,
as claimed in
	the Introduction
--
their standpoint would require explaining,
however,
	how the origin and constancy of $h$
		can follow completely from classical theory
	for radiation,
yet imply
	inherently nonclassical physics
		in other contexts
without involving any other constant
	on a similar fundamental scale as $h$.
Zero point energy and superfluidity theories,
for instance,
have only $h$ to formally represent
	their inherent quantumness in the related equations,
and Schr\"odinger's equation would describe
	a merely classical state space evolution
but for the factor
	$\hbar \equiv h/2\pi$
	(\cf \cite[\S27]{Dirac})
--
with $h$ now shown to be a classical constant,
	akin to Boltzmann's constant $k_B$,
classical explanations for these theories
	must presumably exist and need to be uncovered.
The notion of observer states
	being fundamentally involved in quantum measurements
appears to be new,
	and may be necessary for the exercise
for reasons
	suggested in the Introduction.

\appendix

\section{Atomicity of wavelength intervals}\label{a:lobes}

\begin{lemma}[Wave interval energies] \label{t:genphase} 
	The energy of
		a complete wavelength interval of a periodic wave
	is invariant of
		its initial phase.
	The energy of
		a half-wavelength interval of a sinusoidal wave
	is invariant of
		its initial phase.
\end{lemma}
\proof
	We need to establish,
		in terms of the functional notation of
			Section \ref{s:implied},
	that
	$
	\partial U(2 \pi, \alpha, f) / \partial \alpha = 0
	$
		identically,
	meaning that
		the total energy in the volume of the segment
			remains unchanged
		even as the wave travels through the volume,
	and presents
		different initial phase angles $\alpha$.
	This can be done in two parts,
	first, proving that
		the $\alpha$-derivative of $U(2\pi, \alpha, \sin)$,
			given by equation (\ref{e:phlobe}),
		vanishes,
	and second, that
		the $\alpha$-derivative of the sum $U(2\pi, \alpha, f)$,
			given by equation (\ref{e:fseries}),
		vanishes as well.
	From the elementary calculus of trignometric functions,
	we have
	\begin{equation} \label{e:sineconst}
		\frac{\partial} {\partial \alpha}
			\int_0^{\phi}
			\sin^2 (\theta + \alpha) \, d \theta
	=
		\frac{\partial} {\partial \alpha}
			\int_{\alpha}^{\alpha +\phi}
			\sin^2 \theta \, d \theta
	=
		\frac{\partial} {\partial \alpha}
		\left[
			\frac{\phi}{2}
			-
			\frac{\sin(2 \alpha + 2 \phi) - \sin(2 \alpha)}{4}
		\right]
		=
		0
	\iff
		\phi = n \pi
		,
	\end{equation}
	which yields
		$\partial U(2\pi, \alpha, \sin) / \partial \alpha = 0$
	on setting $n = 2$.
	This proves the second statement in the theorem.

	The second part of this proof involves
		a basic property of a Fourier series expansion,
	that it only contains wavelengths $\lambda_i$
		that are \emph{exact factors} of $\lambda$,
	since each of these components must have the same value
		at each end of the wavelength segment of $f$,
	\ie
		$\lambda/\lambda_i = n$,
	where $n$ is an integer.
	Therefore,
	\begin{equation} \label{e:seriesconst}
		\frac{\partial} {\partial \alpha}
		U(2 \pi, \alpha, f)
		= \sum_{i = 0}^{\infty}
			F_i
			\;
			\frac{\partial} {\partial \alpha}
			U \left(\frac{2 \pi\lambda}{\lambda_i},
				\frac{\alpha\lambda}{\lambda_i},
				\sin
			\right)
		= \sum_{i = 0}^{\infty}
			F_i
			\;
			\frac{\partial} {\partial \alpha}
			U (2 \pi n_i, \alpha n_i, \sin )
		= 0
		\text{~identically.}
	\end{equation}
	It would be noticed that
	this result holds in general only for
		segments of length $\lambda$,
			the full period of the waveform,
	whereas for sinusoidal functions, 
	and indeed
		any alternating functions of $50\%$ duty cycle,
	the constancy would also hold for
		segments of half the wavelength.
	This difference is reflected in the fact that
	equation (\ref{e:sineconst})
		has a null for every $\phi = n \pi$,
	but equation (\ref{e:seriesconst}) involves nulls
		of the component sinusoids only at
			$\phi = 2 \pi n_i$.
\Qed 



\section{\label{a:modes}Density of modes} 

Let $k = 2\pi/\lambda$,
the wavenumber representation for frequency.
Then for $j$-th mode, with $j$ half-wavelength intervals,
	$k_j = j \pi / L$,
	where $L$ = cavity dimension along the wave.
Separation between successive modes is then
	$\delta k = k_{j+1} - k_{j} = \pi/L$.
The number of standing wave modes in the interval $\Delta k$ is then
	$\Delta N = \Delta k / \delta k = L \Delta k / \pi$.
For the three dimensions of physical space,
	$\Delta N(k) = V \Delta^3 k / (2\pi)^3$
or
\begin{equation}
	d N(\mathbf{k}) = \frac{V d^3 \mathbf{k}}{(2\pi)^3} .
\end{equation}
Using $|\mathbf{k}| = \omega/c$,
	transforming to spherical coordinates,
and
	multiplying by $2$ for polarization,
this becomes
\begin{equation}
	d N =
		2 \times \frac
			{V 4 \pi \omega^2 \, d\omega}
			{(2\pi)^3 c^3}
	=
		V . \frac{8 \pi \nu^2 \, d \nu}{c^3}
	,
\end{equation}
yielding the volume density of modes
	used in equation (\ref{e:Planck}),
\begin{equation}
	n(\nu) \, d \nu \equiv
		dN (\nu)
	=
		\frac{8 \pi \nu^2 \, d \nu}{c^3}
	.
\end{equation}


\section{\label{a:hj}Relation to Hamilton-Jacobi theory} 

The following derivation,
	which presumably follows the historical development
		in the early part of the past century, 
is taken directly
	from Goldstein's text
	\cite[10-8]{Goldstein}.
The classical Hamilton-Jacobi theory predicts
	propagating wavefronts in a configuration space,
described by the evolution of 
	Hamilton's principal and characteristic functions 
as (equation 10-142, \ibid)
\begin{equation} \label{e:hjt}
	S(q, P, t) = W(q, P) - Et
\end{equation}
where $S$ and $W$ are
	the principal and characteristic functions, respectively, 
		and $E$ is the energy of the system.
Equation (\ref{e:hjt}) describes
	contours of $S$ moving like waves,
		but the waves are not \apriori sinusoidal.
As in electromagnetic wave theory,
	moving sinusoidal components are obtained
by a trial solution of the form
	$e^{\mathbf{A}(\mathbf{r})\pm
		i (\mathbf{k}\cdot\mathbf{r} \pm \omega t)}$
	(equations 10-152 through 10-155, \ibid)
which then becomes
	an eigenfunction of the wave equation (\ref{e:hjt})
of amplitude
	$a(\mathbf{r}) \equiv e^{\mathbf{A}(r)}$.
By the first principles of integral calculus,
the total solution must be
	a sum or integral of all such eigenfunctions,
\ie
\begin{equation} \label{e:hjFT}
	S(q, P, t)
	=
		\int \;
			a(\mathbf{r})
			\;
			e^{i (\mathbf{k}\cdot\mathbf{r} - \omega t)}
		\; d\omega
	,
\end{equation}
where the domain of $\mathbf{r}$ is the configuration space of
	the coordinates $q$ and momenta $P$.
Equation (\ref{e:hjFT}) simply defines
	the Fourier transformation of $S$.
The correspondence to quantum theory comes from
	the eikonal approximation
\begin{equation} \label{e:eikonal}
	(\nabla L)^2 = n^2
\end{equation}
where
	$L$ is the phase or optical path length,
and
	$n$ signifies a refractive index.
This requires the notion of wave velocity
	$u = E / |\nabla W|$,
and is valid only in
	the limit of slowly changing $n$.
For a single particle,
Hamilton-Jacobi theory yields
\begin{equation} \label{e:gradw}
	(\nabla W)^2 = 2 m (E - V)
\end{equation}
where $V$ is the potential energy
	and $m$, the mass.
The similarity of
	equations (\ref{e:eikonal}) and (\ref{e:gradw})
suggests a correspondence between $W$ and $L$,
	or equivalently between
		$S \equiv W - Et$
	and the total phase of the eigenfunction
		$k(L - ct)$,
implying a proportionality of the particle energy to
	the frequency of the eigenfunction,
		$E = \bar{h} \nu \equiv \bar{h} \omega / 2 \pi$,
where
	$\bar{h}$ is simply a constant of proportionality.
Using this,
the particle motion can be expressed by the evolution equation
\begin{equation} \label{e:pmotion}
	\psi =
		\psi_0
		\;
		e^{i 2 \pi S / \bar{h}}
	,
\end{equation}
which would be a solution of the corresponding
	``Schr\"odinger's equation''
for the particle,
\begin{equation} \label{e:se}
	\frac{\hbar^2}{2m} 
		\nabla^2 \psi
		-
		V \psi
	=
		\frac{\hbar}{i}
		\frac{\partial \psi}{\partial t}
\end{equation}
	with $\hbar = \bar{h}/2\pi$. 
Indeed, in terms of $S$,
	equation (\ref{e:se}) becomes
\begin{equation}
	\left[
		\frac{1}{2m}
		(\nabla S)^2
		+
		V
	\right]
		+
		\frac{\partial S}{\partial t}
	=
		\frac{i \hbar}{2 m}
		\nabla^2 S
\end{equation}
which matches
	the classical Hamilton-Jacobi equation for $S$
iff
	$\hbar \nabla^2 S \ll (\nabla S)^2$,
corresponding to
	the short wavelength limit
at which
	the potential $V$ varies little over a wavelength.
This correspondence is currently interpreted as 
	reducing quantum theory,
		as represented by equation (\ref{e:se}),
	to the classical motion
		given by equation (\ref{e:pmotion})
if $\hbar$ were zero,
	this being Bohr's correspondence principle.
Goldstein paraphrases this perspective,
	which evolved over the first half of the past century,
by pointing out that
	Hamilton could not have discovered Schr\"odinger's equation
because
	lacking ``experimental authority for the jump'',
he had no reason to believe that
	``the value of $h$ was at all different from zero''.

This idea, including Bohr's principle, too is fallacious
because equation (\ref{e:pmotion}),
	$
	\psi = \psi_0 \; e^{i 2 \pi (W - Et)/\bar{h}}
	$,
defines a Fourier component from the decomposition,
	equation (\ref{e:hjFT}),
so that
	$\hbar \equiv \bar{h}/2\pi$
is fundamentally
	\emph{a scale factor for the frequency domain.}
Consequently,
	\emph{zero is the only value not permissible for $h$},
since the decomposition as such would become undefinable
	with this choice.
It thus follows that
	\emph{Hamilton already had mathematical authority
		for $h > 0$}.

The only missing data were the magnitude of $h$,
	which had to be experimentally determined,
and, more importantly,
	the thermodynamic significance presented here.
It is only this significance
	that makes the experimentally determined value,
of approximately
		$6.654 \times 10^{-34}
			~\joule
			~\second
		$,
	special for identifying the Fourier scale factor
		$\bar{h}$ with Planck's constant $h$,
just as a related thermodynamic significance makes
	the experimental value of
		$
		1.3802 \times 10^{-23}
		~\joule
		$
	special for $k_B$.

---------------------------------------------------------------------

\end{document}